\documentclass[trackchanges]{aastex701}

\usepackage{amsmath} 
\usepackage{url}
\usepackage{hyperref}

\begin{document}

\title{The Properties of Non-Potential Magnetic Field Parameters in a Super-Active Region with Complex Structures and Strong Solar Flares\\

\href{https://ui.adsabs.harvard.edu/abs/2025SoPh..300...51L}{ADS: 2025SoPh..300...51L};
\href{https://doi.org/10.1007/s11207-025-02456-6}{DOI}  }

\author[0000-0002-1396-7603,gname=Suo,sname=Liu]{Liu,S.}

\affiliation{National Astronomical Observatories, Chinese Academy of Science, Beijing, 100101, China}
\email[show]{lius@nao.cas.cn}

\author[orcid=0009-0008-2912-1136,gname=Idrees, sname=Shahid]{Shahid Idrees} 

\affiliation{National Astronomical Observatories, Chinese Academy of Science, Beijing, 100101, China}
\email{lius@nao.cas.cn}

\author[gname=Dina, sname=Liu]{Liu, D.} 

\affiliation{Liao Ning University, Shenyang, 110036, China}
\email{}

\author[gname=Shuguang, sname=Zeng]{Zeng, Sh.G.} 

\affiliation{College of Science, China Three Gorges University, Yichang 443002, China}
\email{ }

\begin{abstract}
Solar active regions (ARs), characterized by intense magnetic fields, are prime locations for solar flares. Understanding the properties of these magnetic fields is crucial for predicting and mitigating space weather events. In this study, the non-potential magnetic field parameters of active region NOAA 9077 are investigated; this AR experienced a super-strong X5.7 solar flare.
Using advanced extrapolation techniques, the 3D magnetic field structure from vector magnetograms is obtained from the Solar Magnetic Field Telescope (SMFT) at Huairou Solar Observing Station (HSOS). Then various non-potential parameters are calculated, including current density, shear angle, quasi-separatrix layers (QSLs), twist, and field line helicity.
By analyzing the spatial and temporal distributions of these parameters, we aim to shed light on the relationship between magnetic field properties and solar flare occurrence. Our findings reveal that high twist and complex magnetic field configurations are prevalent before flares, while these features tend to weaken after the eruption. Additionally, we observe decreases in helicity and free energy after the flare, while the free energy peaks approximately 1.5 days prior to the onset of the flare. Furthermore, we investigate the distribution of quasi-separatrix layers and twist, finding high degrees of complexity before flares. Multiple patterns of high current density regions suggest unstable magnetic structures prone to flaring, coinciding with shear angle distribution. Relative field line helicity patterns exhibit distinct characteristics compared to current density, concentrating before flares and diverging afterward. Overall, our results highlight the contrasting nature of current density and relative field line helicity patterns in relation to solar flares, in addition to the aforementioned feature in the set of commonly derived non-potential parameters for this particular event.

\end{abstract}




\section{INTRODUCTION}
The Sun's magnetic field is the driving force behind a variety of solar active phenomena, including flares, filament eruptions, coronal mass ejections, and smaller-scale events. These phenomena are fueled by the immense energy stored in the magnetic field \citep{1995ApJ...451L..83S, 1994ApJ...424..436W, 1996ApJ...456..840T, 2000JGR...10523153F, 2002A&ARv..10..313P, 2012ApJ...748L...6N}.
Understanding the intricate relationship between magnetic fields and solar flares is crucial for predicting and mitigating space weather events. Specific physical parameters derived from magnetic fields have emerged as key indicators for studying the characteristics of both magnetic fields and solar flares \citep{1990SoPh..125..251M, 2001SoPh..204...11D, 2002ApJ...573L.133N, 2008SoPh..248...67L}. Since the energy powering solar flares originates from non-potential magnetic energy, non-potential parameters in active regions have become a focal point in investigating the interplay between the non-potential characteristics of magnetic fields and solar flares \citep{1993SoPh..148..119L, 2002ApJ...573L.133N, 2003ApJ...594.1033N, 2007AdSpR..39.1835L, 2012SoPh..280..165Y}. \citet{2012SoPh..280..165Y} conducted a comprehensive statistical analysis of the non-potentiality of photospheric magnetic fields in solar active regions and their correlation with associated flares. Using vector magnetic field data from the Solar Magnetic Field Telescope (SMFT) at Huairou Solar Observing Station (HSOS), they examined data spanning Solar Cycles 22 and 23. The study revealed a positive correlation between the examined non-potential magnetic field parameters and the flare productivity of active regions.

Non-potential parameters, directly derived from photospheric vector magnetic fields, are essential tools for investigating the intricacies of magnetic fields and solar flares. Two commonly used parameters are vertical current density and shear angle.
Vertical current density, induced by the magnetic field, reflects the curl of the magnetic field and can be calculated using the equation $\textbf{J}=1/{\mu_0}(\nabla\times \textbf{B})$. The shear angle, on the other hand, is the difference between the azimuthal angles of the measured transverse field and the corresponding extrapolated potential fields.
These two parameters are derived directly from the observed vector magnetograms on the photosphere, while other parameters rely on the extrapolated coronal magnetic field in a 3D volume. One such parameter the field line helicity (FLH: \citealt{2018JPlPh..84f7702Y}) differs from the traditional volume helicity, which calculates a single value across the entire active region \citep{1984JFM...147..133B, 1999PPCF...41B.167B}. For the FLH, magnetic field lines are traced, and appropriate vector potentials under specified gauges are established. Subsequently, FLH is computed within the active region for individual field lines, enabling the investigation of magnetic helicity distribution based on field line tracing. Both helicity and twist quantify the degree of twisting in a magnetic field, with helicity being more closely associated with magnetic field energy. Twist is inherently linked to the topological structure of field lines and is determined by the ratio of the magnetic field relevant components. In contrast, FLH exhibits a stronger dependence on the intensity of individual magnetic field components compared to twist, suggesting that FLH encapsulates a richer set of information about magnetic field energy.

Solar flares, energetic eruptions on the Sun's surface, pose a significant threat to Earth's technology and environment. Understanding the underlying magnetic field structures of active regions is crucial for predicting and mitigating these events.
This study primarily focuses on investigating non-potential parameters derived from the spatial magnetic field to explore the properties of active region magnetic fields and flares. Concurrently, non-potential parameters calculated directly from the photospheric magnetic field are also included and analyzed. To obtain the coronal magnetic field of an active region, the MHD-relaxation method is employed to utilize the conservation element/solution element (MHD-CESE) numerical scheme \citep{2013ApJ...769..144J}, where the vector magnetic field captured by SMFT/HSOS serves as the bottom boundary condition for extrapolation. Based on the extrapolated coronal magnetic field, quasi-separatrix layers (QSL, \citealt{2007ApJ...660..863T}; \citealt{2016ApJ...818..148L}), twist \citep{2016ApJ...818..148L}, and field line helicity (FLH, \citealt{2018JPlPh..84f7702Y}; \citealt{2019A&A...624A..51M}; \citealt{2021A&A...649A.107M}) are calculated to gain insights into the magnetic field properties. Furthermore, the current density and shear angle that are highly representative of non-potential parameters are also investigated.

The article is organized as follows: Section \ref{sec:obser} outlines the observational data and methodological approach, Section \ref{sec:results} presents the results of our analysis, and Section \ref{sec:disc} summarizes our findings and discusses their implications.

\section{Observations and Methods}\label{sec:obser}

NOAA AR 9077, a prominent active region during Solar Cycle 23, is the subject of our research. From July 11 to 15, 2000, the Solar Magnetic Field Telescope (SMFT) at the Huairou Solar Observing Station (HSOS), National Astronomical Observatories (NAOC), Chinese Academy of Sciences (CAS), conducted meticulous observations of the vector magnetic field of the region. On July 14, 2000, this active region produced a powerful X5.7 solar flare near the disk center, known as the second Bastille Day flare of Solar Cycle 23. This event significantly impacted Earth and remains a pivotal event in solar cycle research.
Previous studies using SMFT data \citep{2001A&A...372.1019L, 2001SoPh..204...11D, 2016AdSpR..57.1468L} identified high shear magnetic field configurations and successive flux emergence as key factors contributing to this X5.7 flare. Additionally, \citet{2016AdSpR..57.1468L} analyzed changes in general magnetic field parameters like the height of field lines, the force-free factor, and free energy using 3D magnetic field extrapolations. In contrast, this study leverages SMFT vector magnetic field data and the sophisticated MHD-CESE extrapolation method to focus on advanced, high-quality non-potential magnetic parameters associated with the spatial magnetic field structures. This approach enables a deeper analysis of the magnetic field structure and its role in driving the powerful flare.

The Solar Magnetic Field Telescope (SMFT) is a 35-centimeter refracting telescope designed to observe the Sun's magnetic field. It employs the magnetically sensitive Fe {\sc i} $\lambda$5234.19\,\AA{} spectral line, with a Lande $g$ factor of 1.5, to measure the splitting of energy levels induced by the magnetic field. This line has an equivalent width of approximately 0.33\,\AA{}. The birefringent filter of SMFT, with a narrow bandwidth of 0.15\,\AA{}, allows for precise measurements of both longitudinal and transverse magnetic fields. By shifting the central wavelength of the filter by -0.075\,\AA{}, the telescope can isolate specific spectral components to determine the magnetic field components \citep{1986AcASn..27..173A}.

Vector magnetograms are reconstructed from four narrowband images capturing Stokes parameters  $I$, $Q$, $U$ and $V$. Specifically, $V$ represents the difference between left and right circularly polarized light, while $Q$ and $U$ characterize linear polarization in different directions. The intensity of the light is given by the $I$ parameter. To derive the magnetic field components from these Stokes parameters, a linear calibration is necessary, which is valid under the weak-field approximation \citep{1989ApJ...343..920J, 1991ApJ...372..694J}. The magnetic field components can be derived from the corresponding Stokes parameters based on the following equations:
\begin{equation}
B_{L}=C_{L}V ,
\end{equation}
\begin{equation}
B_{T}=C_{T}(Q^{2}+U^{2})^{1/4},
\end{equation}
\begin{equation}
\theta=arctan(\dfrac{B_{L}}{B_{T}}),
\end{equation}
\begin{equation}
\phi=\dfrac{1}{2}arctan(\dfrac{U}{Q}),
\end{equation}
where $B_{L}$ and $B_{T}$ denote the line-of-sight (LOS) and transverse components, respectively, of the photospheric magnetic field. The angle $\theta$ signifies the inclination between the vector magnetic field and the line of sight direction, whereas the angle $\phi$  indicates the azimuth of the field. Whereas, $C_{L}$ and $C_{T}$ act as calibration coefficients for the longitudinal and transverse magnetic fields, respectively, and they are paramount in determining the accuracy of the vector magnetic field. To make $C_{L}$ and $C_{T}$ more accurate a series of experiments are needed. For example, scanning data at multiple wavelength positions are collected, and by fitting the theoretical and actual spectral line profiles, the calibration coefficients are obtained. Then the suitable $C_{L}$ and $C_{T}$ are used for the vector magnetic field calibration.
In this study, $C_{L}$ and $C_{T}$ are set to 8381 G and 6790 G, as reported by \citet{2004ChJAA...4..365S}.
Following standard data processing at the Huairou Solar Observing Station (HSOS), our observations achieve a spatial resolution of roughly 2 arcsec/pixel. This translates to highly detailed images with a noise level of 20 G for the longitudinal component and 150 G for the transverse component (both at the 3 $\sigma$ level). The data is publicly available for further exploration at https://sun10.bao.ac.cn/hsos\_data/magnetogram.
A key challenge in solar magnetic field measurements is the 180-degree ambiguity in the transverse field component. This means the telescope cannot directly distinguish between opposite magnetic field directions. We address this by employing the minimum energy method. This technique simultaneously minimizes both the divergence of the magnetic field ($\nabla\cdot\textbf{\emph{B}}$) and the electric current density ($\textbf{\emph{J}}=\nabla\times\textbf{\emph{B}}$) within the Sun's atmosphere. This minimization is achieved using a sophisticated algorithm known as simulated annealing \citep{1994SoPh..155..235M, 2006SoPh..237..267M}.

The observational data from the Huairou Solar Observing Station (HSOS) has been extensively utilized in various research areas, including magnetic field studies, investigations of magnetic field non-potentiality, and magnetic field extrapolation for magnetic helicity calculations \citep{2018JPlPh..84f7702Y, 2022ApJ...929..122W}.

\begin{figure}
   \centering
   \includegraphics[width=\linewidth]{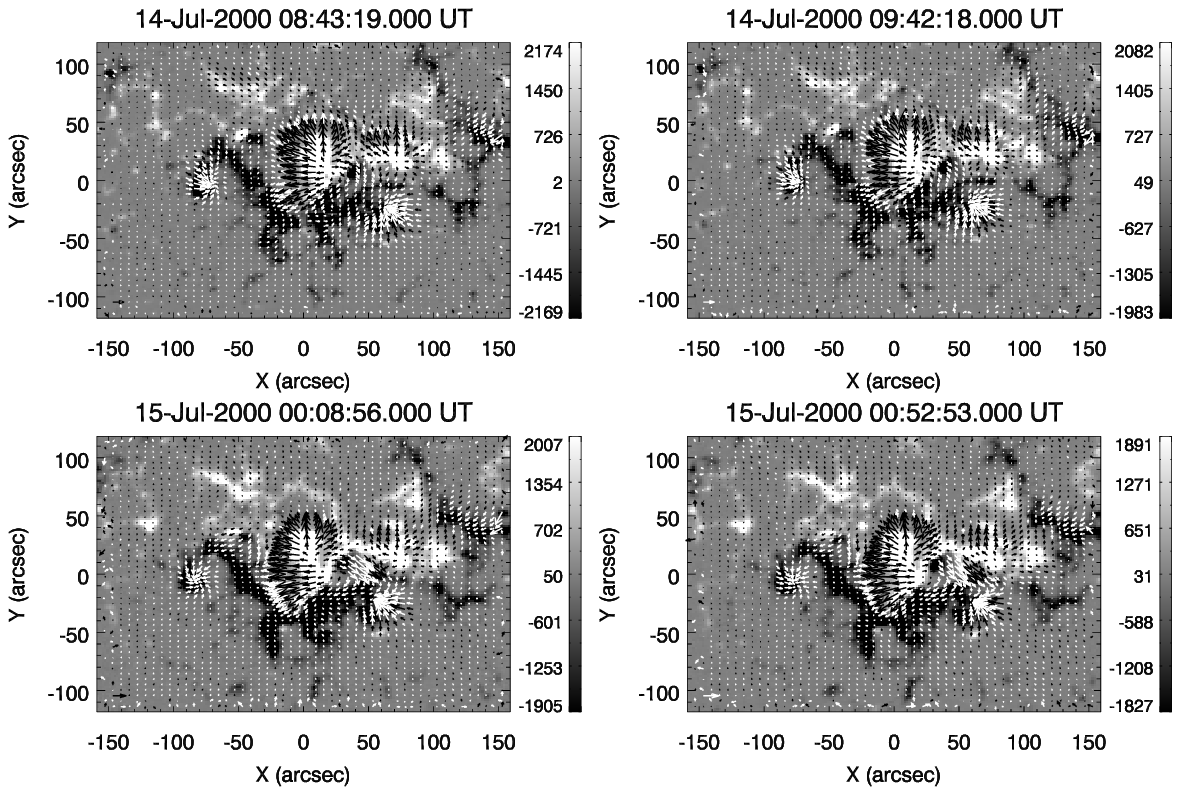}
   \caption{The magnetograms of the AR 9077 captured by SMFT with a field of view of 5.23$'$ $\times$ 3.63$'$ before (upper row) and after (bottom row) the flare, with the observed times labelled in each panel.
Here the grayscale image represents the longitudinal magnetic field, with the magnitude of the field indicated by the colorbar. The arrow directions and lengths indicate the directions and strengths of the transverse magnetic field.}
   \label{boundaryambig}
\end{figure}

Figure \ref{boundaryambig} presents a series of vector magnetograms captured by the Solar Magnetic Field Telescope (SMFT) before and after the powerful X5.7 flare that erupted on July 14, 2000, at 10:03 UT. These images offer a detailed view of the active region NOAA 9077, covering a field of view of 5.23$'$ $\times$ 3.63$'$ with a spatial resolution of $\approx$ 2 arcsec/pixel.
The grayscale images represent the longitudinal magnetic field, while the arrows superimposed on the images indicate the direction and strength of the transverse magnetic field. The active region exhibits a complex, super-active structure, characterized by the prominent $\beta\gamma\delta/\beta\gamma\delta$ magnetic configurations, which is a preferred magnetic structure for flare production.

\begin{figure}
   \centering
   \includegraphics[width=\linewidth]{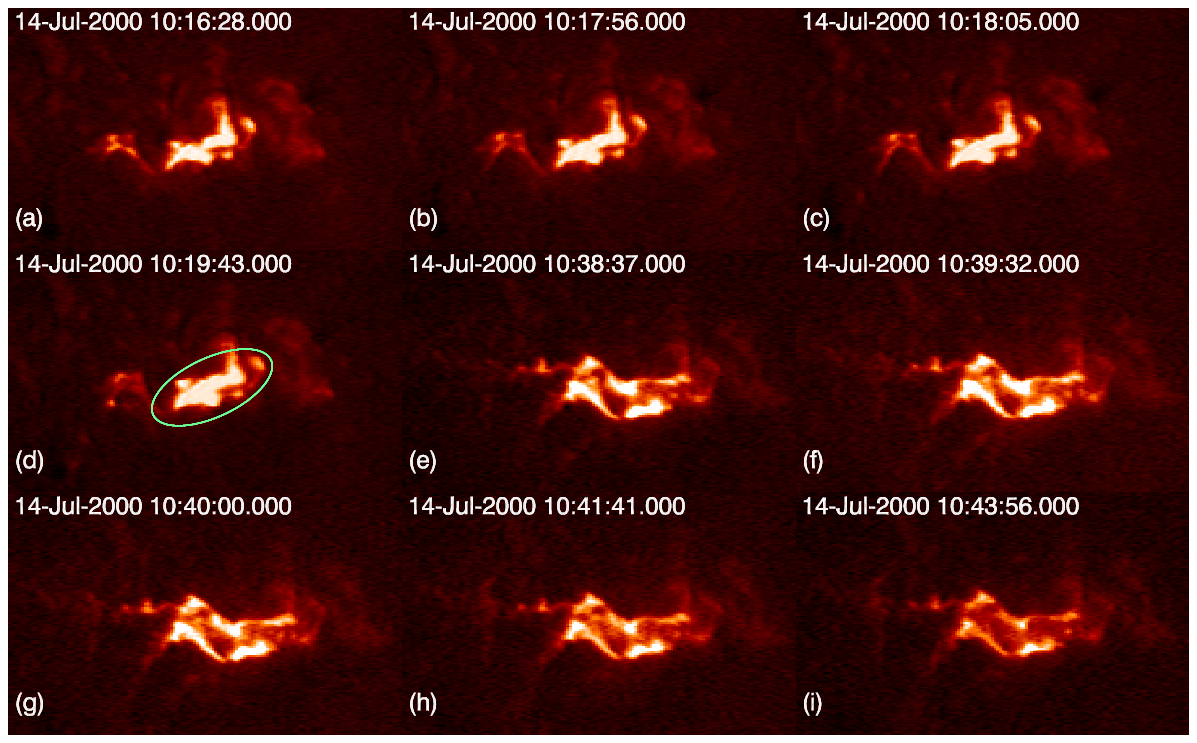}
   \caption{The evolution of H$\alpha$ emission for NOAA AR 9077 during the X5.7 flare observed by SMFT. The green oval outlines the main flare ribbon during the time of maximum flare intensity.}
   \label{hrha9077}
\end{figure}

Figure \ref{hrha9077} presents a sequence of H$\alpha$ images captured by the Solar Magnetic Field Telescope (SMFT) during the X5.7 flare event. This flare unfolded over a period of approximately 40 minutes, beginning at 10:03 UT, reaching its peak intensity at 10:24 UT, and finally subsiding at 10:43 UT.
A striking feature of this flare is the emergence of a prominent airplane-shaped primary flare ribbon, highlighted by the green oval in panel d. This ribbon marks the region where the initial magnetic reconnection event likely occurred. Subsequently, the flare evolved into a more typical dual-ribbon configuration.
The complex magnetic topology of the active region, characterized by the X-shaped reconnection site (the sub-region marked by the green oval can be regarded as an X-shaped region of magnetic reconnection, where the initial magnetic reconnection event occurs), provided the necessary conditions for this powerful eruption.

In this study, the extrapolation strategy is to apply the MHD-CESE non-linear force-free field (NLFFF) method, to a time-series of vector magnetogram. This method is a type of MHD relaxation technique utilizing the conservation-element/solution-element (CESE) solver, to derive the three-dimensional (3D) coronal magnetic field in a near-force-free state. The CESE method numerically solves the complete MHD equations, while the bottom boundary condition is set analogous to the stress-and-relax approach, incrementally modifying the transverse field to align with the magnetogram. Meanwhile, the remaining boundaries of the computational box are governed by nonreflecting boundary conditions, an adaptive grid technology is also used in this method. This method has been extensively utilized for analyzing magnetic field topologies using real solar magnetic field data \citep{2011ApJ...727..101J, 2013ApJ...769..144J, 2017ApJ...842..119D, 2023ApJ...958...90H}. The extrapolated fields ought to approximately fulfill the solutions of the force-free equations (($\nabla\times\textbf{B})\times\textbf{B}=0, \nabla\cdot\textbf{B}=0$) thus necessitating the verification of the extent of force-free. To assess the validity of the extrapolated field, the force-free criterion  $\sigma_J$  (equation \ref{sigmaj}) and the divergence-freeness criterion $f$ (equation \ref{ff}) are employed and given below, as outlined by \citet{2000ApJ...540.1150W} and \citet{2009ApJ...696.1780D}:
\begin{equation}\label{sigmaj}
\sigma_J=\dfrac{\sum_iJ_i\sigma_i}{\sum_iJ_i},
\end{equation}
and
\begin{equation}\label{sigmaj1}
\sigma_i=sin\theta_i=\dfrac{|\textbf{J}\times\textbf{B}|_i}{J_iB_i},
\end{equation}
\begin{equation}\label{ff}
f_i=\dfrac{\int_{\Delta S_i}\textbf{B}\cdot\textbf{dS}}{\int_{\Delta S_i}|\textbf{B}|dS}\approx\dfrac{(\nabla\cdot\textbf{B})_i\Delta V_i}{B_iA_i}.
\end{equation}

\begin{figure}
   \centering
   \includegraphics[width=\linewidth]{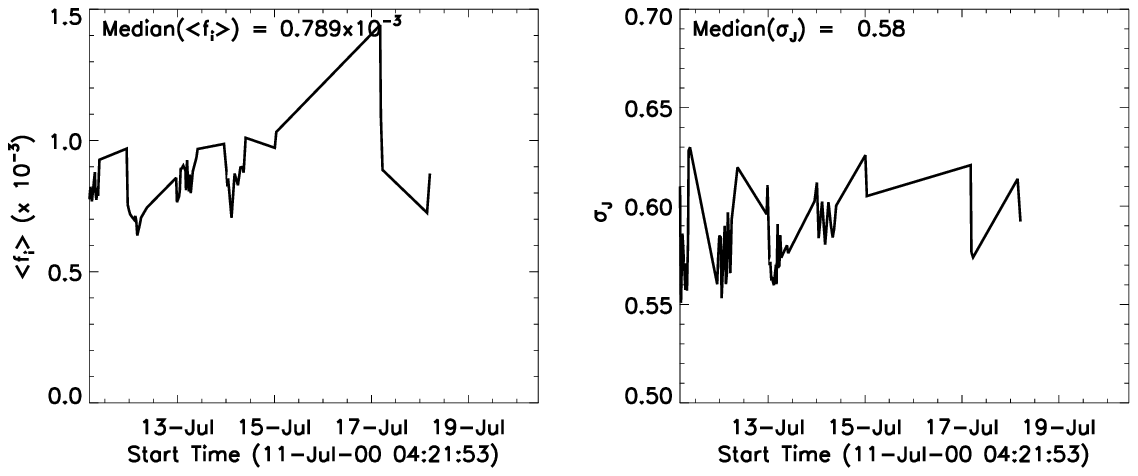}
   \caption{The values of $<$$f_i$$>$ and $\sigma_{J}$ calculated from the series of 3D extrapolated magnetic field for AR 9077, observed by SMFT from July 11 to 19 2000. The median values are 0.789 $\times10^{-3}$
and 0.58, respectively.}
   \label{ffsigmaj}
\end{figure}

Figure \ref{ffsigmaj} illustrates the distribution of $<$$f_i$$>$ and  $\sigma_J$ , for the entire extrapolated field of NOAA 9077, derived from 57 magnetograms obtained by the SMFT. The values of $<$$f_i$$>$ predominantly range between 0.6 x 10$^{-3}$ and 1.0 x 10$^{-3}$, with a median value of 0.789 x 10$^{-3}$. Only one value exceeds 1.0 x 10$^{-3}$. Meanwhile, $\sigma_J$ values cluster around 0.55 to 0.65, with a median of 0.584. These findings align well with previous studies by \citet{2000ApJ...540.1150W} and \citet{2014Ap&SS.351..409L}, who reported  $\sigma_J$ values around 0.405 and $<$$f_i$$>$ values around 0.5 x 10$^{-4}$. The reasonable agreement of these parameters suggests the validity of our subsequent analysis.

The primary advantage of obtaining a coronal magnetic field model for the active region lies in enabling our ability to investigate the intricate topological properties of the magnetic field. By tracking the 3D distribution of magnetic field lines, we can delve into the detailed morphology and quantify key parameters: the squashing factor (QSL), twist, and field line helicity (FLH).
To calculate QSLs and twist, we employed the methods and codes developed by \citet{2016ApJ...818..148L} and \citet{2022ApJ...937...26Z}. These techniques, optimized for 3D data cubes, leverage graphics processing unit (GPU) acceleration, and a step-size adaptive fourth-order Runge-Kutta scheme for efficient field line tracking. For FLH calculations, we adopt the methods and codes pioneered by \citet{2018JPlPh..84f7702Y}, which involve determining appropriate vector potentials under specific gauges and calculating the relative field-line helicity (RFLH) within the active region. To quantify FLH, we track magnetic field lines individually based on the extrapolated field and computed the helicity in each fied lines. However, instead of considering the entire active region volume, we focus on individual field lines, as defined by thin magnetic flux tubes ($D_i$), as shown in Equation 8,
\begin{equation}\label{flh0}
h(D_i)=\int_{D_i}\bf{A}\cdot\bf{B}d^3x
\end{equation}

\begin{figure}
   \centering
   \includegraphics[width=\linewidth]{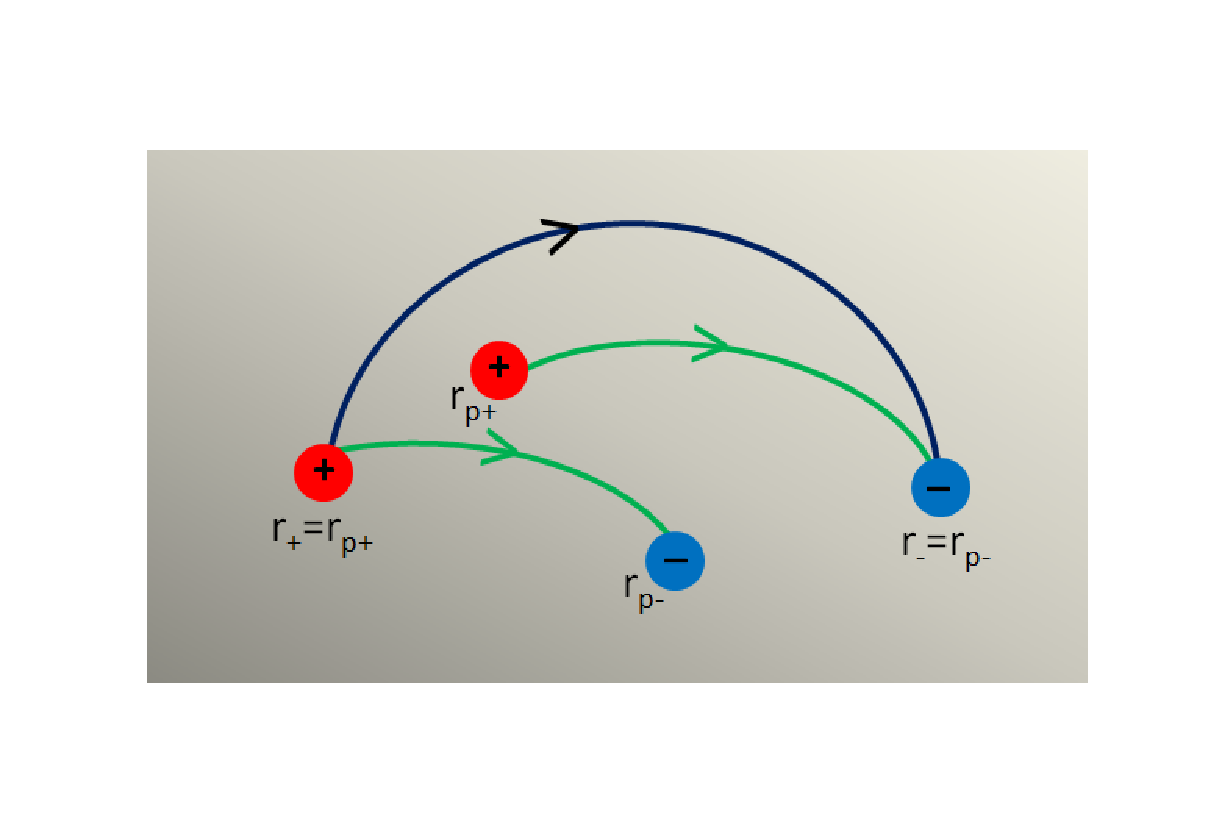}
   \caption{The illustration of spatial magnetic field lines in the definition of the relative field line helicity for equations \ref{flh1}-\ref{flh3}. The blue line indicates one closed field line of the extrapolated field, where the positive and negative roots are labbeled by red and blue circles. The green lines are the corresponding field lines from the reference potential field.}
   \label{flhfig}
\end{figure}

Similar to helicity, the field line helicity (FLH) has to be gauge invariant. To ensure a consistent definition of relative field-line helicity (RFLH), a specific gauge condition must be imposed. By computing the vector potential for the the extrapolated field and a reference field ($\textbf{A}_p$) under this gauge condition, RFLH ($H_R$) can be calculated using Equations \ref{flh1}-\ref{flh3}, as illustrated in Figure \ref{flhfig}. Where  $r+$/$r-$ and $r_{p{^+}}$/$r_{p{^-}}$  labelled the roots of corresponding field lines \citep{2018JPlPh..84f7702Y}. Significantly, for FLH this approach has been successfully utilized for analyzing RFLH and current in real magnetic field observations \citep{2021A&A...649A.107M}.

\begin{equation}\label{flh1}
H^+_R=\int_{r_+}^{r_-}\bf{A}\cdot d\bf{l}-\int_{r_+}^{r_{p{^-}}}\bf{A_p}\cdot d\bf{l_p}
\end{equation}
\begin{equation}\label{flh2}
H^-_R=\int_{r_+}^{r_-}\bf{A}\cdot d\bf{l}-\int_{r_{p{^+}}}^{r_-}\bf{A_p}\cdot d\bf{l_p}
\end{equation}
\begin{equation}\label{flh3}
H_R=\dfrac{1}{2}(H^+_R+H^-_R)
\end{equation}

\section{Results}\label{sec:results}

\begin{figure}
   \centering
   \includegraphics[width=\linewidth]{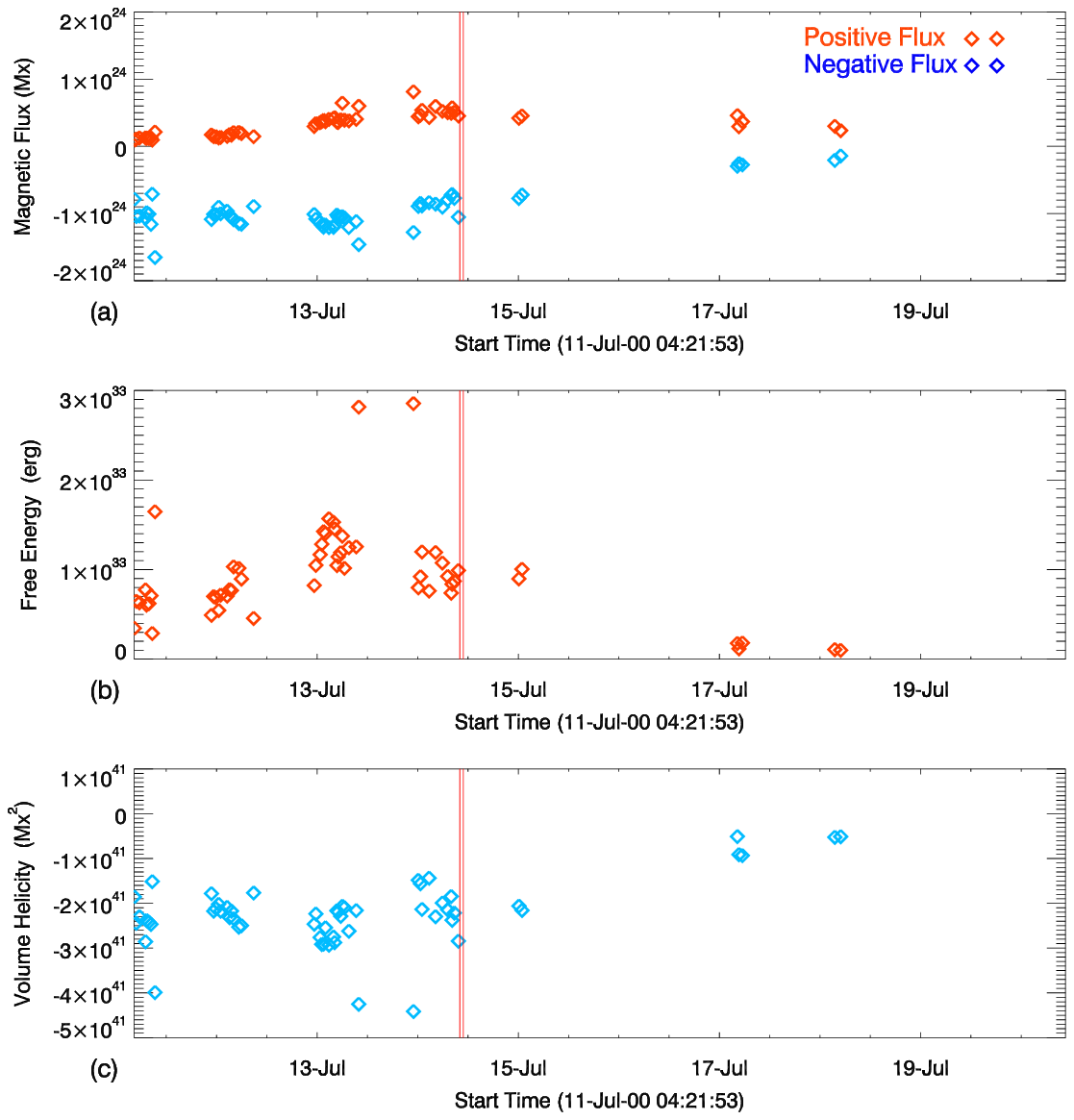}
   \caption{The evolution of (a) magnetic flux, (b) free energy, and (c) volume helicity in the AR 9077 as observed by SMFT. Positive and negative flux are shown in red and blue, respectively. Two vertical red lines mark the start and end time of X5.7-class flare event.}
   \label{pfnf.freee.hr.cese}
\end{figure}

Figure \ref{pfnf.freee.hr.cese} provides a comprehensive overview of the temporal evolution of magnetic fluxes, categorized by polarity (positive: red diamonds, negative: blue diamonds), alongside free energy and volume magnetic helicity. The time frame analyzed extends from July 11 to 18, 2000. Each panel highlights a significant event: an X5.7 flare marked by two vertical red lines. These data were acquired using the Solar Magnetic Field Telescope (SMFT) and derived from coronal magnetic field extrapolations using the MHD-CESE method. The calculation of free magnetic energy adheres to Equation \ref{eqfree}, while the relative volume magnetic helicity is determined using Equation \ref{eqvhr}. Here, \textbf{B} represents the extrapolated coronal magnetic field, and \textbf{B$_p$} denotes the potential field with identical photosphere boundary conditions. The quantities \textbf{A} and \textbf{A$_p$} correspond to the vector potentials of \textbf{B} and \textbf{B$_p$}, respectively.

This figure offers a holistic view of the magnetic flux, free energy, and helicity evolution. A striking observation is the substantial post-flare decrease in both positive and negative magnetic fluxes. Interestingly, the peak in free magnetic energy precedes the major flare by approximately 1.5 days (on July 13), followed by a gradual decline without immediate fluctuations around the flare time. However, a significant drop is evident between July 18 and 19. Throughout the lifespan of the AR, consistently negative relative volume magnetic helicity is observed. A notable post-flare decrease in volume helicity is also evident.

\begin{equation}\label{eqfree}
E_{free}=E-E_p=\int\dfrac{B^2}{8\pi}dV-\int\dfrac{B_{p}^2}{8\pi}dV
\end{equation}
\begin{equation}\label{eqvhr}
H_R=\int_{V}(\bf{A}+\bf{A}_p)\cdot(\bf{B}-\bf{B}_p)dV
\end{equation}

\begin{figure}
   \centering
   \includegraphics[width=\linewidth]{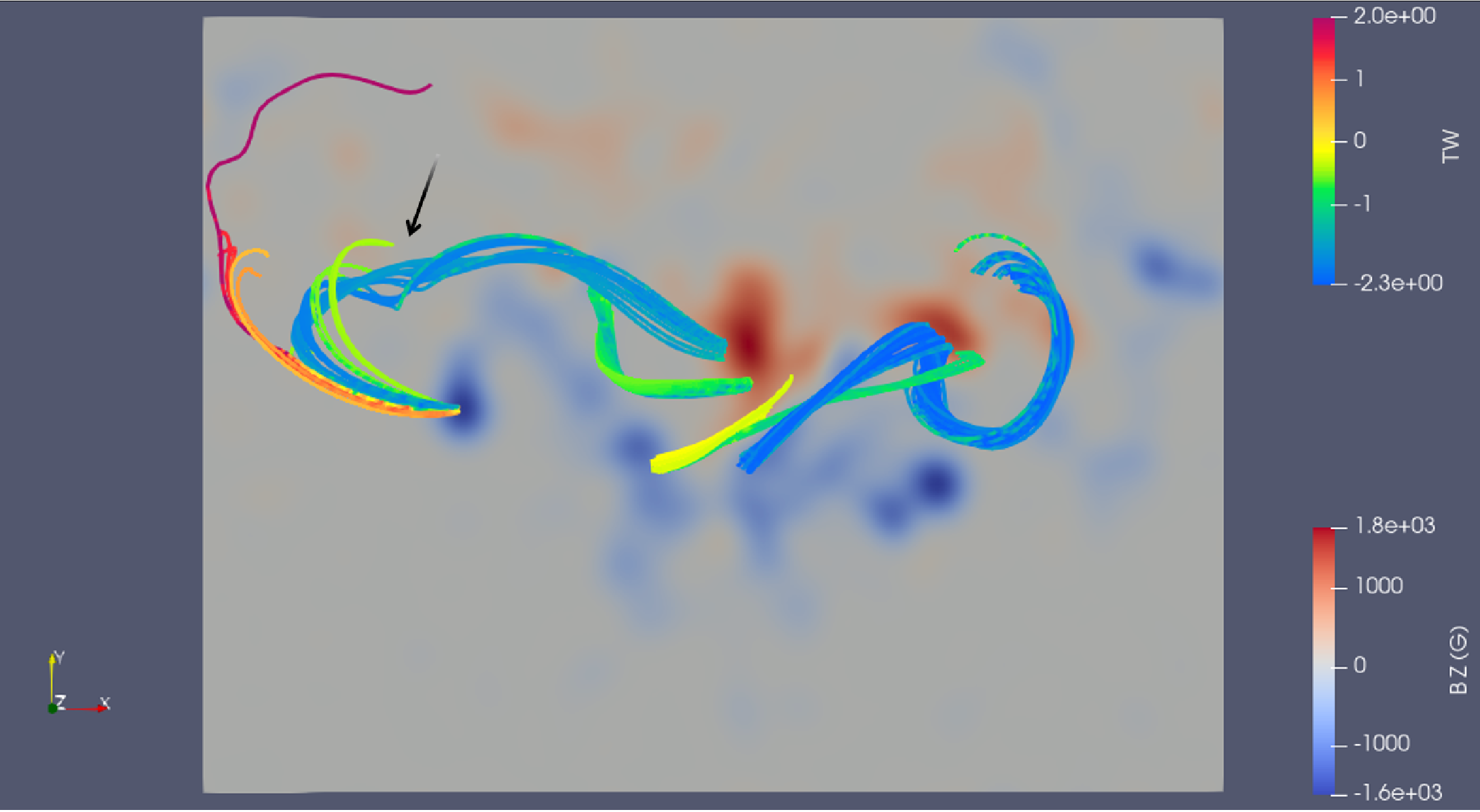}
   \caption{Magnetic field lines in the active region before the X5.7 flare on 14 July 2000 09:42:18 UT. The colorbar labeled B$_{Z}$ shows the intensity z-component of the magnetic field in Gauss, the colorbar labeled by TW indicates the twist number of a field line in dimensionless unit.}
   \label{linedistr1}
\end{figure}

\begin{figure}
   \centering
   \includegraphics[width=\linewidth]{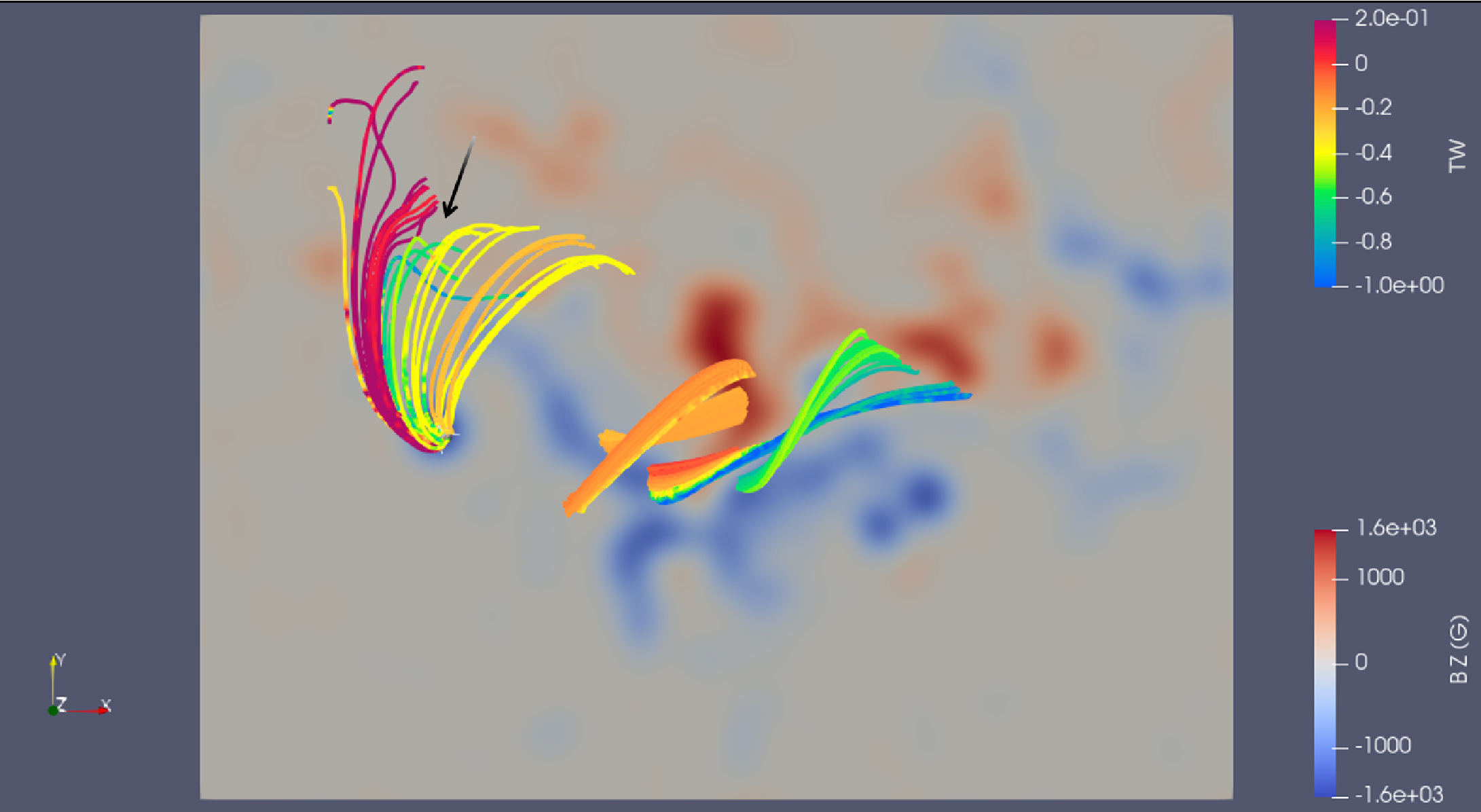}
   \caption{The same as Figure \ref{linedistr1}, but for a time after the occurrence of the X5.7 flare at 15 July 2000 00:08:56 UT.}
   \label{linedistr2}
\end{figure}

Figures \ref{linedistr1} and \ref{linedistr2} provide a visual representation of the magnetic field lines before(14 July 2000 09:42:18 UT) and after (15 July 2000 00:08:56 UT) the X5.7 flare, respectively, where the field line is painted by the value of its individual twist number. These figures offer insights into the topological evolution of the extrapolated magnetic field within the active region.
Prior to the flare, the magnetic field exhibited a complex topology with prominent, highly twisted magnetic field lines entwining each other. These twisted flux tubes closely correspond to the elongated structures observed in H${\alpha}$ images during the flare, as depicted in panels a-d of Figure 2. Notably, the magnetic field lines in the upper left corner of the active region (indicated by a \textbf{black} arrow in each figure) display a higher degree of twist before the flare compared to the post-flare configuration. Additionally, significant changes in connectivity were observed, such as the disappearance of long, highly twisted loops connecting the central region to the negative polarity region, which means that the topological shift towards a potential field configuration is consistent with the observed dual-band flare ribbon in H${\alpha}$ images.

\begin{figure}
   \centering
   \includegraphics[width=\linewidth]{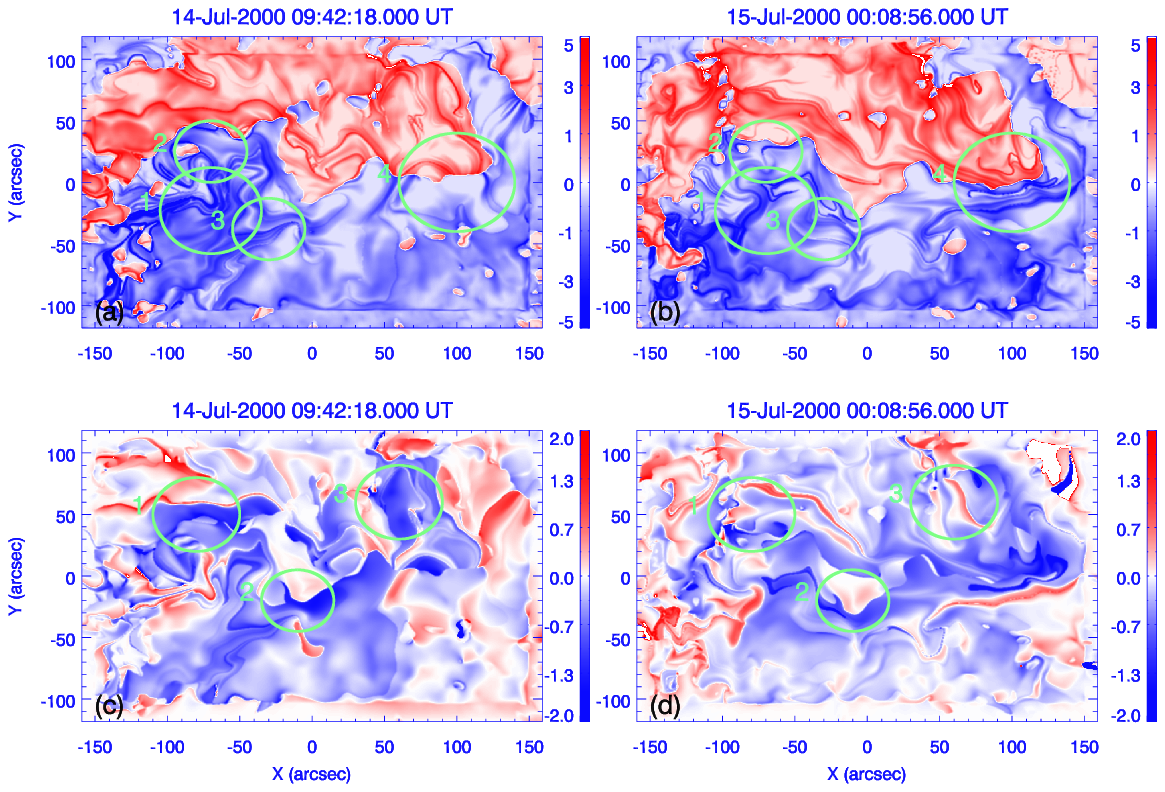}
   \caption{(a) and (b): The distributions of the logarithmic squashing factor (Q) on the photosphere before and after the occurrence of the X5.7 flare. (c) and (d): The distributions of magnetic field twist on the photosphere, also before and after the X5.7 flare. In each panel, the green circles outline the region for comparison purposes, highlighting changes before and after the flare.}
   \label{qsltwist4950_cese}
\end{figure}

This active region presents a complex magnetic configuration, allowing for a detailed quantitative analysis. Figure \ref{qsltwist4950_cese} a and b illustrate the squashing factor Q, a metric derived from photospheric footpoints, which indicates the locations of quasi-separatirx layers (QSLs) a??regions of high Q often associated with intricate 3D magnetic structures. The twist value (Figure \ref{qsltwist4950_cese}(c) and (d)) further underscores the non-potential nature of the magnetic field.
A notable pre-flare characteristic is the increased complexity of QSLs within the green circled regions (1, 2, and 3). This suggests a more intricate magnetic connectivity before the flare, which may have facilitated energy release. After the flare,, the QSLs appear to simplify, possibly indicating a more coherent magnetic configuration.
Similarly, the twist distribution exhibits a higher concentration of strong twists prior to the flare, especially within the green circled regions. This aligns with the observation of large, highly twisted magnetic field lines (Figure  \ref{linedistr1}) in regions 2 and 3, which weaken or disappear after the flare,.
These findings provide compelling evidence for the role of magnetic reconnection in driving solar flares. The complex pre-flare magnetic topology, characterized by intricate QSLs and strong twists, likely provides the necessary conditions for energy release. After the flare, simplification of the magnetic field suggests a relaxation of the stressed magnetic configuration, consistent with the dissipation of energy through reconnection.

\begin{figure}
   \centering
   \includegraphics[width=\linewidth]{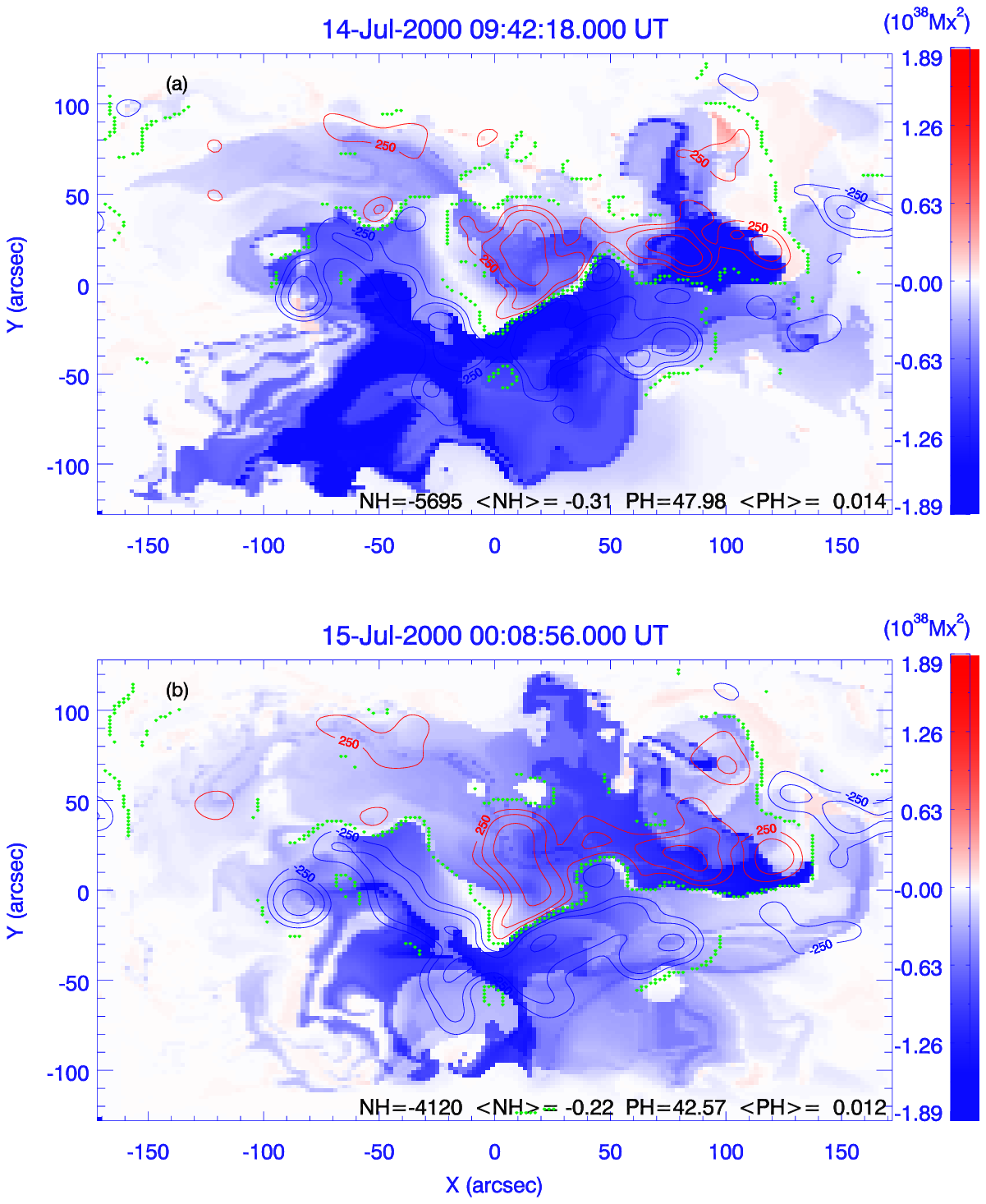}
   \caption{The distribution of RFLH on the photosphere before (a) and after (b) the occurrence of an X5.7 flare. NH and $<$NH$>$ labled are the total and mean of negative RFLH for this region, while PH and $<$PH$>$ are total and mean of positive RFLH in unit of 10$^{38}$Mx$^2$. The \textbf{green} contours indicate the magnetic neutral line, while red and blue contours outline the positive and negative line-of-sight magnetic fields, respectively.}
   \label{rflh4950}
\end{figure}

A method developed by \citet{2018JPlPh..84f7702Y} is used to calculate relative field line helicity (RFLH). RFLH effectively quantifies the helicity contained within individual magnetic field lines. Each pixel on the photospheric boundary represents the helicity of the corresponding field line (some field line end at the side or top boundary). The sum of all these pixel values equals the total volume helicity.
Figure \ref{rflh4950} presents the RFLH distribution on the photosphere before and after the X5.7 flare. The yellow contour marks the magnetic inversion line, while red and blue contours indicate positive and negative line-of-sight magnetic fields, respectively. Before and after the flare, negative RFLH dominates the entire active region, aligning with the volume magnetic helicity findings (see Figure \ref{rflh4950}c).
It is found that the high values of RFLH are concentrated primarily before the flare onset, but after the flare, the distribution of RFLH exhibits greater diversity. Furthermore, these high values of RFLH tend to be located near the inversion line.
Utilizing the distributions of RFLH in the active region, both the total and mean values of positive and negative RFLH are calculated and labeled as NH/PH and $<$NH$>$/$<$PH$>$, respectively.
Notably, after the flare, the amplitudes of both positive and negative helicities diminish, both in terms of total and mean values. Specifically, for negative RFLH, the total value drops from -5695 to -4120$\times 10^{38}$ Mx$^2$, while the mean values decline from -0.31 to -0.22$\times 10^{38}$ Mx$^2$. On the other hand, for positive RFLH, the total value decreases slightly from 47 to 42 $\times 10^{38}$ Mx$^2$, and the mean values dip from 0.014 to 0.012 $\times 10^{38}$ Mx$^2$. Notably, the variations in negative RFLH constitute a significant portion of the overall RFLH variances, and these changes closely correlate with the majority of changes in the non-potentiality of the magnetic field.

\begin{figure}
   \centering
   \includegraphics[width=\linewidth]{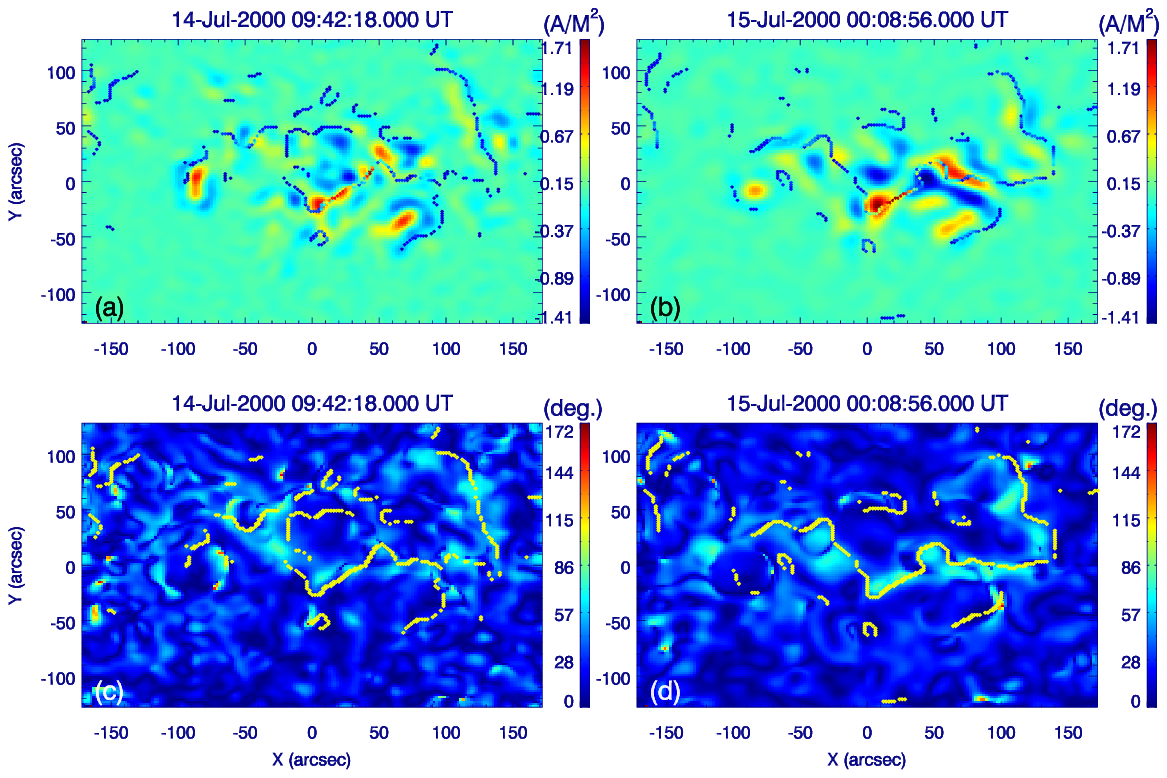}
   \caption{(a) and (b): The distributions of the vertical current density $J_z$ (A/m$^2$) observed on the photosphere before and after the occurrence of the X5.7 flare. (c) and (d): The distributions of shear angle (degrees) on the photosphere before and after the flare. In each panel, the magnetic neutral lines are outlined by blue/yellow dotted lines.}
   \label{currentphi4950}
\end{figure}

Figure \ref{currentphi4950} provides insights into the local non-potential properties of the active region before and after the flare by visualizing the vertical current density ($J_{z}$) and shear angle. $J_{z}$ is derived from the curl of the vector magnetic field at the z-direction, and it represents the strength of electric currents flowing vertically within the solar atmosphere. Whereas, the shear angle quantifies the deviation of the observed magnetic field from a potential field, which is a hypothetical field configuration free of currents.
Before the flare, $J_{z}$ exhibits a more diverse distribution, with high values scattered across the active region. Similarly, the shear angle shows a more dispersed distribution, with significant deviations from the potential field. After the flare, the distribution of $J_{z}$ becomes more concentrated, suggesting a reorganization of the currents. The shear angle distribution also becomes more focused, with a reduction in the overall level of non-potentiality. Both $J_{z}$ and the shear angle tend to exhibit higher values near magnetic inversion lines, both before and after the flare. This indicates that these regions are associated with stronger currents and deviations from the potential field. The more concentrated distribution of $J_{z}$ after the flare aligns with the observed two-ribbon flare morphology in the H$\alpha$ images (Figure 2). This suggests a connection between the reorganization of currents and the flare energy release.
The observed changes in the distribution of $J_{z}$ and shear angle before and after the flare provide valuable clues about the underlying physical processes. The concentration of current and shear near the inversion lines suggests that these regions play a crucial role in energy storage and release during solar flares. The post-flare reorganization of currents, as indicated by the more concentrated $J_{z}$  distribution, may be related to the dissipation of magnetic energy and the formation of the flare ribbons.

\section{Discussions and Conclusions}\label{sec:disc}

This study delves into the complex interplay between solar flares and non-potential magnetic field parameters within super-active regions. By employing advanced techniques, we reconstruct the three-dimensional magnetic field configuration of NOAA AR 9077 using Solar Magnetic Field Telescope (SMFT) observations and MHD-CESE extrapolation method. This enables a meticulous analysis of critical magnetic field parameters, including current density, shear angle, quasi-separatrix layers, twist, and field line helicity.
Our analysis reveals strong correlations between these non-potential parameters and the occurrence of solar flares. The spatial distribution and temporal evolution of these parameters exhibit distinct patterns that precede and accompany solar eruptions. These findings underscore the crucial role of non-potential magnetic field configurations in driving solar flare activity.
By gaining a deeper understanding of these parameters, we can significantly improve our ability to predict and model solar flare events. This knowledge is not only scientifically valuable but also has practical implications for space weather forecasting, which is essential for safeguarding technological infrastructure and human activities in space.
In conclusion, our study provides compelling evidence for the importance of non-potential magnetic field parameters in solar flare initiation and evolution. Further research in this area will contribute to the development of more accurate and reliable solar flare prediction models.

Our analysis reveals a significant decrease in magnetic helicity and free energy following the X5.7 flare. Notably, free energy reaches a maximum approximately 1.5 days prior to the flare. Additionally, a detailed examination of the spatial distribution of these parameters within the active region provides crucial insights into the complex magnetic field dynamics leading to the flare.
By tracing magnetic field lines using a 3D magnetic field model, we observe a substantial, highly twisted magnetic flux configuration before the flare. This twisted structure weakens or dissipates post-flare, suggesting a relaxation of the magnetic field. Furthermore, certain field lines with rotational structures either lose their intensity or undergo changes in their connectivity patterns. Overall, the magnetic field transitions towards a more potential configuration with reduced twist and simpler structures after the flare. This transition aligns with the observed flare characteristics in $H{\alpha}$ images.
Before the flare, the distributions of quasi-separatrix layers (QSLs) and twist exhibit a high degree of complexity, indicating the presence of multiple unstable magnetic systems. After the flare, this distribution becomes simpler, and the topological changes align with those observed in QSL and twist features. For instance, the weakening of twisted in specific regions corresponds to the weakening of twist field lines or changes in their connectivity.
Regions of high current density ($J_{z}$) may imply the presence of unstable magnetic structures, which are consistent with the distribution of shear angles. In contrast, the distribution patterns of relative field line helicity (RFLH) differ from those of $J_{z}$ . While RFLH is concentrated before the flare, it tends to disperse afterward. These findings are consistent with previous studies, such as \citeauthor{2021A&A...649A.107M}, (\citeyear{2021A&A...649A.107M}).

While this study provides valuable insights into the role of non-potential magnetic field parameters in solar flare activity, significant questions remain unanswered. The precise mechanisms behind the accumulation and release of magnetic energy in complex active regions are still not fully understood. Future research should leverage advanced numerical models and comprehensive observational data to gain a deeper understanding of these processes.
The findings presented in this study highlight the importance of non-potential magnetic field parameters in comprehending solar flare activity within super-active regions. By analyzing these parameters, we can uncover crucial information about the relationship between magnetic fields and flares, as well as identify potential indicators for predicting and modeling solar flare events. Continued research in this field will contribute to a more comprehensive understanding of the intricate dynamics of the solar magnetic field and its role in driving solar activity.
\begin{acknowledgments}
We sincerely thank the editor and referee for their thorough review of our manuscript and for providing valuable feedback, which has significantly enhanced the quality of our paper. We are also grateful to the editor for their meticulous language editing, which has greatly improved the clarity and readability of the manuscript.
\end{acknowledgments}



\bibliographystyle{aasjournalv7}
\bibliography{liu.bib}



\end{document}